\documentclass[aps,pre,one,11pt,showpacs]{revtex4}
\usepackage{epsfig}
\usepackage{graphicx}
\usepackage{amsmath}
\usepackage{amsfonts}
\usepackage{amssymb}%
\begin{document}
\title{Self-assembly of binary nanoparticle dispersions: from square arrays and
stripe phases to colloidal corrals.}
\author{Carlos I. Mendoza \footnote{Author to whom correspondence should be
addressed. Tel: +52(55)56224644, Fax: +52(55)56161201, email:
cmendoza@iim.unam.mx} and Erasmo Batta}
\affiliation{Instituto de Investigaciones en Materiales, Universidad Nacional Aut\'{o}noma
de M\'{e}xico, Apdo. Postal 70-360, 04510 M\'{e}xico, D.F., Mexico}

\begin{abstract}
The generation of nanoscale square and stripe patterns is of major
technological importance since they are compatible with industry-standard
electronic circuitry. Recently, a blend of diblock copolymer interacting via
hydrogen-bonding was shown to self-assemble in square arrays. Motivated by
those experiments we study, using Monte Carlo simulations, the pattern
formation in a two-dimensional binary mixture of colloidal particles
interacting via isotropic core-corona potentials. We find a rich variety of
patterns that can be grouped mainly in aggregates that self-assemble in
regular square lattices or in alternate strips. Other morphologies observed
include colloidal corrals that are potentially useful as surface templating
agents. This work shows the unexpected versatility of this simple model to
produce a variety of patterns with high technological potential.

\end{abstract}
\maketitle

%\pacs{}

It has been a challenge in nanotechnology to produce highly ordered structures
by controlling the position of nanoparticles over an extended length scale
\cite{madueno}; it is not easy to achieve this goal using direct micro- and
nano-fabrication since such processes are prohibitively expensive and
time-consuming below a certain length scale. As a result, the search of
particles on the mesoscopic scales that self-organize\ into potentially useful
structures by virtue of their mutual interactions is extremely important
\cite{rechtsman1}, \cite{rechtsman2}. The ability of soft-matter systems to
self-assemble in a surprisingly large variety of periodic arrangements has
lead to use it as building blocks of bottom-up nanofabrication processes. This
is a large and rapidly growing field of tremendous technological potential and
fundamental interest that has been feeded by the continuing progress in the
manipulation of the interaction potentials between nanoparticles
\cite{min}-\cite{tang}. In particular, the self-assembly of
nanometer-length-scale patterns in two dimensions is currently of interest as
a method for improving throughput and resolution in nanolithography
\cite{bita}; however, traditional self-assembling approaches based on block
copolymer lithography spontaneously yield nanometer-sized hexagonal structures
that are incompatible with the square arrangements used in industry-standard
circuits\cite{tang}. That is the reason why among the targeted morphologies,
stripes and square arrays are particularly important for the manufacture of
microelectronic components since they would enable simplified addressability
and circuit interconnection in integrated circuit nanotechnology. There have
been theoretical efforts to tailor interparticle interactions to spontaneously
produce target many-particle configurations for single component systems
\cite{rechtsman1}, \cite{rechtsman2} but the optimized potentials found are
difficult to fit by available interactions. Recently, an experimental
technique based in a blend of diblock copolymers was developed to produce
nanoparticles that assemble into square arrays \cite{tang}. The blend consists
of two kinds of diblock copolymers, A-B and B'-C, being the A and C blocks
mutually repulsive and incompatible with blocks B and B'. Block B contains
small numbers of groups that form hydrogen bonds to complementary groups in
block B' (Fig. 1a). The attractive interactions between complementary hydrogen
bonds suppress macrophase separation in favor of microphase separation,
thereby producing large-scale assembly of nanoscale features. By controlling
the amount of hydrogen bonding units, the molecular weights and compositions
of the block copolymers, diverse families of ordered structures are achieved,
including square arrays of cylinders that are the result of the competition
between the different interactions (see Fig. 3 of Ref. \cite{tang}). Inspired
by these experiments, we propose a simple model system that reproduces such
structures as a result of a binary mixture of nanoparticles interacting via
isotropic short ranged competing potentials. The model we propose treat the
interactions between the block copolymers in an effective way that takes into
consideration the entropy of the brush-like coronas surrounding each cylinder
via a soft repulsive shoulder and the hydrogen bonding via a square well
attraction. Such simple pair potentials are often used to describe effective
interactions among substances with supramolecular architecture
\cite{malescio3},\cite{denton}. We consider a two-dimensional (2D) assembly of
particles consisting of a hard core surrounded by a soft corona which is
repulsive for particles of the same species and attractive otherwise. The
simplicity of the model allows to predict, based on geometrical arguments,
under which circumstances a given family of structures is obtained. Among the
rich variety of patterns encountered, we can highlight the formation of square
arrays and stripe phases for equimolar mixtures, and the formation of
colloidal corrals for asymmetric mixtures. These results suggest a strategy
for producing a range of self-assembled structures which could be exploited
for use as templating or directing agents in materials syntheses. Although we
are primarily interested in the crystalline structures at low temperatures, we
find that at higher temperatures the system presents a transition from an
isotropic fluid phase to a disordered fluid-like stripe phase.

Our model is not limited to the above mentioned diblock copolymer blend since
core-corona architectures are also present in numerous physical systems such
as dendritic polymers, hyper-branched star polymers, etc. Among these are, for
instance, colloidal particles with block-copolymers grafted to their surface
where self-consistent field calculations lead to effective interactions that
can be modeled by a square-shoulder potential \cite{norizoe}. Such
interactions can be controlled by adjusting the length, species, the grafting
density of the grafted polymers, the quality of the solvent, the density and
location of the hydrogen bonding, etc.\ Numerical simulations have shown that
single component softened-core repulsive potentials may give rise to strip
phases \cite{norizoe}-\cite{fornleitner} and periodic structures that are
explained in terms of the competing interactions between the hard core and the
soft shoulder. Our aim is to study the influence that the introduction of a
second component has in the domain formation. The second component interacts
attractively with the first one in order to suppress macroscopic phase
separation. This attractive interaction models the hydrogen bond attraction
between B and B' segments of the diblock copolymer system of Ref. \cite{tang}.
When this system forms cylindrical micelles, as shown schematically in Fig.
1b, then our model treat them as nanoparticles with a core-corona architecture
where the hard core represents the blocks A and C and the soft corona models
the effective interactions between the blocks B and B'. Therefore, our system
consists of a binary mixture of particles interacting through a radially
symmetric pair potential composed by an impenetrable core of diameter
$\sigma_{0}$ with an adjacent square shoulder with range $\lambda\sigma_{0}$
(Fig. 1c), \emph{i.e.},%
\begin{equation}
\phi(r)=\left\{
\begin{array}
[c]{cc}%
\infty, & r\leq\sigma_{0},\\
\pm\epsilon, & \sigma_{0}<r<\lambda\sigma_{0},\\
0, & r\geq\lambda\sigma_{0},
\end{array}
\right.  \label{phi}%
\end{equation}
$r$ being the pair distance. Particles of the same component interact through
the repulsive shoulder $\epsilon$ whereas particles of different species
interact through the attractive potential well of depth $-\epsilon$. The model
intend to represent the interactions of the block copolymers of Ref.
\cite{tang} as long as they form cylindrical micelles (Fig 1b). At large
distances, the particles do not overlap and the interaction vanishes,\ the
repulsive shoulder models the steric repulsion between blocks B or between
blocks B' due to the overlap of the brush-like coronas, and the attractive
well models the hydrogen bonding between blocks B and B'. Finally, at small
separations penetration of the compact cores is very unfavorable and gives
rise to the hard-core repulsion. The simple functional form of the interaction
potential not only capture the essential features of colloidal particles with
core-corona architecture, it also offers many computational advantages and
allows to understand using simple geometrical considerations the system's
self-assembly strategy \cite{pauschenwein2}. Standard Monte Carlo simulations
based on the canonical ensemble (NVT simulations) in a square box of side $L$
with periodic boundary conditions have been carried out using the Metropolis
algorithm. We have used $\sigma_{0}$ and $\epsilon$ as length and energy
units, respectively, and we have studied the pattern formation dependence on
$\lambda$, reduced temperature $T^{\ast}=k_{B}T/\epsilon$, where $k_{B}$ is
the Boltzmann's constant; reduced number density $\rho^{\ast}=N\sigma_{0}%
^{2}/L^{2}$, and relative amount of each species $x=N_{2}/N_{1}$, with $N_{i}$
the number of particles of species $i$ and $N=N_{1}+N_{2}$ the total number of
particles. Simulations are performed with $N=1000$ particles (runs with
$N=4000$ were done for the largest value of $\lambda$). In all cases, the
system is first disordered at high temperature and then brought from $T^{\ast
}=1$ ($T^{\ast}=10$ for the largest value of $\lambda$) to the final
temperature $T^{\ast}=0.1$ through an accurate annealing procedure with steps
of $0.01$ ($0.1$ for the largest value of $\lambda$). An equilibration cycle
consisted, for each temperature, of at least $2\times10^{6}$ MC steps, each
one representing one trial displacement of each particle, on average. At every
simulation step a particle is picked at random and given a uniform random
trial displacement within a radius of $0.5\sigma_{0}$.

A variety of interesting structural features of this simple model at low
temperatures are in evidence in Fig. 2, where a few of them are exhibited.
Panel (a) shows a representative example of the spatial configuration that the
system adopts for $\lambda=1.5$, $\rho^{\ast}=0.5$, and $x=1$. We observe that
the system self-assemblies forming an aggregate made of an alternating square
array of particles with a lattice parameter determined by the range of the
soft coronas (inset Fig. 1a). The system adopts this array in order to
maximize the number of favorable overlaps between particles of different
species avoiding at the same time unfavorable overlaps between particles of
the same type. Thus, this simple model reproduces qualitatively well the main
characteristic of the block copolymer system of Ref. \cite{tang}, i. e., the
fundamentally and technologically important result that this system
self-assemblies forming square arrays. Unlike systems with only one component
\cite{malescio2}, the ratio between the hard and soft cores is critical even
if they are comparable to each other since different patterns may appear by
just changing the value of $\lambda$. In Fig. 2b we show the structures
obtained for the same set of parameters as in Fig. 2a except that the value of
$\lambda$ has been increased to $\lambda=2$. The particles now form aggregates
in which the particles prefer to align forming intercalated stripes of
alternating species. Each stripe consists of only one type of particles and
try to avoid overlap between the coronas of neighboring stripes of the same
species, thus, the shoulder width act as spacer between neighboring stripes
(see inset of Fig. 2b). Increasing even more the range of the soft corona, the
system develops a completely different strategy to form stable configurations
giving rise to the appearance of new patterns. This is shown in Fig. 2c where
a pattern formed by a square array of tetramers develops if $\lambda$ is
increased to $\lambda=3$. Each square tetramer is formed by four close packed
particles of the same species and the separation between the tetramers is
dictated by the condition that the coronas of each particle overlaps at most
four coronas of particles of the same species, as shown in the inset of Fig.
2c where we have highlighted the coronas that settle the spacing between
tetramers. Two lines of defects, one near the upper side and the second near
the right side of the simulation cell are in evidence. They are originated in
the periodic boundary conditions; effectively, the lattice of tetramers extend
over the entire length of the simulation cell and join up with their periodic
images. Since this length is not commensurable with the periodicity of the
lattice the defect lines appear. A few vacancies scattered along the lattice
that could not be annealed out are also present, but this is to be expected
for kinetic reasons. For very large values of $\lambda$ ($\lambda=10$ in Fig.
2e and f), larger hexagonal close-packed aggregates of particles of the same
species are formed and this aggregates in turn self assemble in square arrays
[panel (e)] or thick stripes [panel (f)] formed similarly of hexagonal
close-packed particles (inset Figs. 2e and f). We can think of the aggregates
as larger interacting particles with the internal structure only changing the
effective inter-aggregate interactions \cite{glaser}, thus the system try to
arrange particles so that the shape of the cluster becomes as circular as
possible. This, in turn, guarantees that the underlying structure is close to
the energetically most favorable hexagonal lattice \cite{fornleitner}. Within
the lane structure (Fig. 2f) the system follows a similar strategy consisting
in first optimize the packing inside a stripe, leading to hexagonal particle
arrangements inside each lane. Nevertheless, due to the increasing complexity
of the inner structure of the patterns, a simple energetic explanation in
terms of overlapping coronas is very difficult.

In our model, configurational energy is essentially determined by geometry;
thus, simple geometrical considerations can be used to predict the parameters
at which a given pattern is expected to be adopted at zero temperature. For
values of the range of the soft coronas such that $\lambda\geq\sqrt{2}$ the
lattice constant $a$ at $T^{\ast}=0$ is determined by the condition that the
soft coronas of the nearest neighbors of the same species just touch as shown
in the inset of Fig. 2a. Therefore, the lattice parameter takes the value
$a=\lambda\sigma_{0}$ and the density of the aggregate is $\rho
_{\text{squares}}^{\ast}=2/\lambda^{2}$. On the other hand, if $\lambda
<\sqrt{2}$ then the particle hard cores just touch forming a close packed
square array whose lattice constant $a=\sqrt{2}\sigma_{0}$ is determined by
the hard-core diameter $\sigma_{0}$, and the density of the aggregate is
$\rho_{\text{squares}}^{\ast}=1$. The energy per particle of this array is
$u=-4\epsilon$ which is the number of overlaps between the corona of a given
particle and its first neighbors times the energy of each overlap. The square
array has the characteristic that no overlap between the coronas of particles
of the same component takes place. However, if the density is so high that the
coronas of particles of the same species have not enough space to accommodate
without overlapping, then the system adopts a different strategy in order to
minimize its energy; the formation of lanes provides the energetically best
solution. This is achieved (if $\lambda\geq2$ ) when the soft corona of a
given particle just touch the ones of the second nearest neighbors of the same
stripe determining the separation between the particles in the same stripe. On
the other hand, if $\lambda<2$, neighboring particles of the same stripe are
in direct contact, forming a one-dimensional close-packed arrangement \ As a
result, in both cases, the particles form a centered rectangular arrangement
as shown in the inset of Fig. 2b, with the lattice parameter in the direction
of the stripe given by $a=\lambda\sigma_{0}/2$ if $\lambda\geq2$ or
$a=\sigma_{0}$ if $\lambda<2$. The lattice parameter perpendicular to the
stripes is given by the distance between adjacent stripes of the same species
which are separated by a distance $b=\lambda\sigma_{0}$ if $\lambda\geq
\sqrt{3}$ or $b=\sqrt{2}\sigma_{0}$ if $\lambda<\sqrt{3}$. The densities are,
correspondingly, $\rho_{\text{stripes}}^{\ast}=4/\lambda^{2}$ if $\lambda
\geq2$, $\rho_{\text{stripes}}^{\ast}=2/\lambda$ if $\sqrt{3}\leq\lambda<2$,
and $\rho_{\text{stripes}}^{\ast}=2/\sqrt{3}\simeq1.1547$ if $\lambda<\sqrt
{3}$; the last situation corresponds to an hexagonal close packing. Using
similar geometrical considerations we obtain the energy per particle in each
case, $u=-6\epsilon$ if $\lambda\geq\sqrt{3}$ or $u=-2\epsilon$ if
$\lambda<\sqrt{3}$. Finally, for the square array of tetramers the lattice
parameter is $a=\sigma_{0}/\sqrt{2}\left(  1+\sqrt{2\lambda^{2}-1}\right)  $
if $\lambda\geq\sqrt{5}$ and the corresponding density is $\rho
_{\text{tetramers}}^{\ast}=16\left(  1+\sqrt{2\lambda^{2}-1}\right)  ^{-2}$.
When $\lambda<\sqrt{5}$ the particles of the adjacent tetramers touch their
hard cores forming a square close packing and therefore $\rho
_{\text{tetramers}}^{\ast}=1$. The energy per particle of this array is given
by $u=-8\epsilon$ if $\lambda\geq\sqrt{5}$, $u=-4\epsilon$ if $2\leq
\lambda<\sqrt{5}$, or $u=0$ if $\lambda<2$.

By comparing to each other the energies of the lattices found previously, it
is possible to construct a phase diagram at zero-temperature which allows to
predict the structures adopted for a given set of parameters. Regions of the
phase diagram where minimum energy configurations at $T^{\ast}=0$ appear are
shown in Fig. 3 for the three simplest structures obtained in the simulations.
Here we see, for instance, that stripes are preferred over squares at larger
densities and for a larger range of the coronas, and that tetramers will be
preferred at even larger values of $\lambda$. The previous results, obtained
using only geometrical arguments, agree with the structures obtained in
Fig.2a, b, and c; nonetheless, we do not expect an exact correspondence
between the phase diagram at $T^{\ast}=0$ with finite temperature simulations
since in that case entropy plays an important role. This is put in evidence in
Fig. 2d which shows the square array adopted in a system with the same
parameters as in Fig. 2b except that the density is reduced to $\rho^{\ast
}=0.3$. According to the phase diagram, for this set of parameters one expects
the formation of stripes at $T^{\ast}=0$; however, since the square array is
less dense, then the particles have more room to vibrate around their
equilibrium positions which in turn means that the contribution to the free
energy $F=U-TS$ due to the entropy associated with the thermal motion is more
important for the square array than for the stripes and therefore the system
may adopt the square array. Thus, polymorphic transitions are expected under
cooling, specially for parameters near the border of the different regions,
i.e., $\lambda\simeq\sqrt{3}$ and $\sqrt{5}$.

Until now we have focused only in the minimum energy configurations, however,
energy considerations gives no information about the mechanical stability of
the structure under the action of external agents \cite{rechtsman1}. The
stability of an hexagonal phase (or an fcc phase in 3D) is often attributed to
the optimality of the close-packed lattice \cite{glaser}. On the other hand,
for the square array with $\lambda\geq\sqrt{2}$ the touching coronas of the
particles belonging to the same species provide mechanical stability to the
structure; however, this is not the case for shorter ranges of the coronas and
the structure is unstable with respect to shear.

Here we are primarily interested in the crystalline configurations that the
system may adopt at low temperatures. Nonetheless, the behavior of the system
at higher temperatures is also very interesting since under cooling other
structures may preempt the crystalline lattice, as previously discussed. In
fact, we have found that the original isotropic fluid phase may first suffer a
transformation to a disordered fluid-like strip phase. This is exhibited in
Fig. 4 where snapshots taken at two temperatures along the annealing isochore
are shown. Starting with an isotropic fluid phase, as the temperature is
decreased the system first shows the formation of a low-temperature fluid
phase [panel (a)] consisting of a disordered stripe structure which under
further cooling undergoes an abrupt transition at a given temperature form the
disordered stripe pattern to a crystalline square array of tetramers [panel
(b)]. For the chosen parameters, the array of tetramers appears to coexist
with a stripped structure in which part of the configurational disorder
typical of high temperatures remains.

The complexity of the patterns grows with increasing asymmetry in the
stoichiometry of the sample. Two representative configurations for
non-equimolar mixtures are shown in Fig. 5. In panel (a) we observe a phase
separation with a region consisting of a high-density blend of the two species
and a region consisting of a low-density phase of the majority component.
Lanes similar to those obtained in the equimolar mixture are observed in the
high-density component. In general we find that patterns similar to those
found for the equimolar mixture appear in the high-density phase. The pure
component may also form patterns in the surrounding regions. As expected, the
structures found in this pure phase are also similar to those found in a
single-component square-shoulder system \cite{malescio1},\cite{malescio2}. A
different behavior is observed in Fig. 5b. Here, there is no macroscopic phase
separation; instead, the system exhibits an ordered hexagonal structure
consisting of clusters of the minority species corralled by particles of the
majority component that form an open network structure. It has been proposed
that these morphologies can be employed for the fabrication of nanostructures
\cite{madueno}. For example, after selectively removing one of the blocks, the
remaining pattern can be transferred into a functional material \cite{bita}%
,\cite{mansky}-\cite{jung}. Similar nanoparticle corrals have recently been
obtained in network-self-assembled monolayer hybrid systems \cite{madueno} and
are of particular interest as surface templates because they contain cavities
that can be filled by a variety of guest molecules. This kind of open network
is particularly flexible since the pore size can be controlled by changing the
range of the corona and the asymmetry of the mixture.

It is known that computer simulations risk to be trapped in local energetic
minima due to the rough and complex energy landscape \cite{fornleitner}; thus,
in order to circumvent this problem we have followed a very accurate simulated
annealing procedure since for sufficiently slow cooling down of the system one
expect that the particles accommodate to its minimum energy configurations
when temperature vanishes \cite{jagla}. Nonetheless, alternative approaches
could also be of great help to predict the complete set of ordered equilibrium
structures of the system; one of these approaches consists in applying genetic
algorithm search strategies to find in a systematic way minimum energy
configurations \cite{pauschenwein1}. Such algorithms have proven to be a
reliable way to predict the ordered equilibrium structures for different
systems, including monolayers of binary dipolar mixtures \cite{fornleitner2}.

In addition to illustrate the complexity of structures that can arise from the
simple model presented in this paper, the system is possible to be carried out
experimentally thanks to the recent advances to control colloidal and
nanoparticle interactions \cite{tang}. From the huge variety of different
ordered equilibrium structures developed in the system, here we have shown
just a few of them that were chosen due to their possible practical relevance
and because some of them have already been observed experimentally
\cite{tang}. The large versatility of the model indicates that such systems
can be adapted to a wide variety of technological applications ranging from
the directed growth of molecular wires or carbon nanotubes \cite{camp} to
nanolithography and nanoelectricity \cite{malescio1} or even as templates for
metal deposition.

\qquad{\LARGE Acknowledgements}

This work was supported in part by Grant DGAPA IN-107607.

\newpage

\begin{figure}[ptb]
\includegraphics[width=8.5cm]{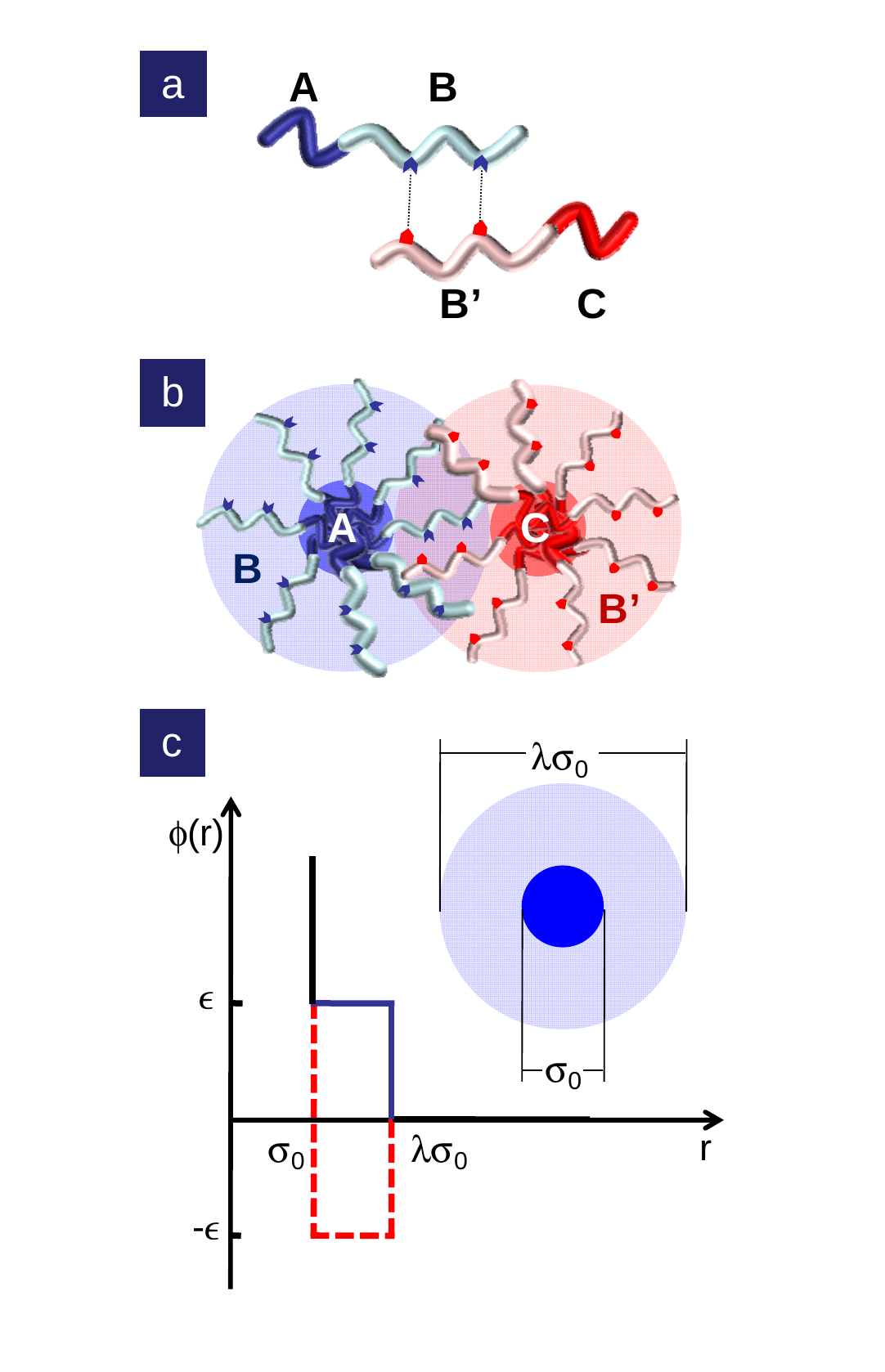}
\caption{Description of the model. (a) Two kinds of diblock copolymer, A-B and B'-C. Block B contains small numbers of groups (blue) that form hydrogen bonds to complementary groups (red) in block B'. (b) Schematic representation of the blend of block copolymer with hydrogen bond interactions forming cylindrical
micelles (top view). (c) The interaction between a pair of cylindrical
micelles is modeled by a hard-core soft-corona potential. If the micelles are
of the same component then the corona is a square shoulder otherwise the
corona is a square well.}%
\label{fig.1}%
\end{figure}

\begin{figure}[ptb]
\includegraphics[width=8.5cm]{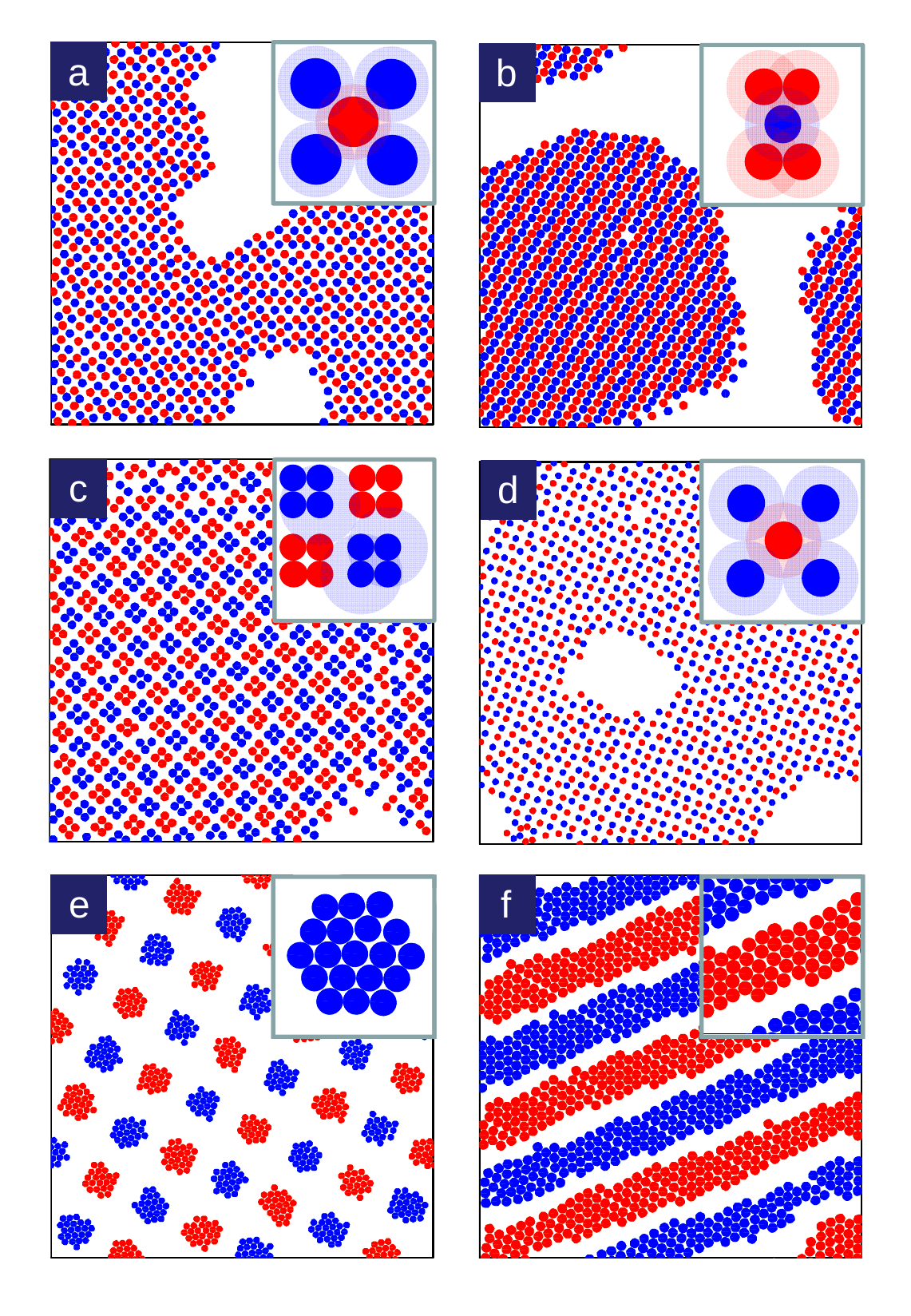}
\caption{Spatial arrangements of the system at temperature $T^{\ast}=0.1$ for
an equimolar mixture $x=1$. (a) Square array formed in a system with
$\lambda=1.5$ and $\rho^{\ast}=0.5$, (b) stripe phase obtained for $\lambda=2$
and $\rho^{\ast}=0.5$, (c) a square array of tetramers appears for
$\lambda=2.5$ and $\rho^{\ast}=0.5$, (d) square array obtained for $\lambda=2$
and $\rho^{\ast}=0.3$. Comparing panels (b) and (d) we see that at the lower
density shown the system aggregates in square arrays whilst at the larger
density the system prefers to form stripes of alternating species. For very
large values of $\lambda$, larger aggregates of hexagonal close-packed
particles of the same species are formed. These aggregates in turn
self-assemble in (e) square arrays for $\lambda=10$ and $\rho^{\ast}=0.3$, and
(f) thick stripes for $\lambda=10$ and $\rho^{\ast}=0.65$.}%
\label{fig.2}%
\end{figure}

\begin{figure}[ptb]
\includegraphics[width=8.5cm]{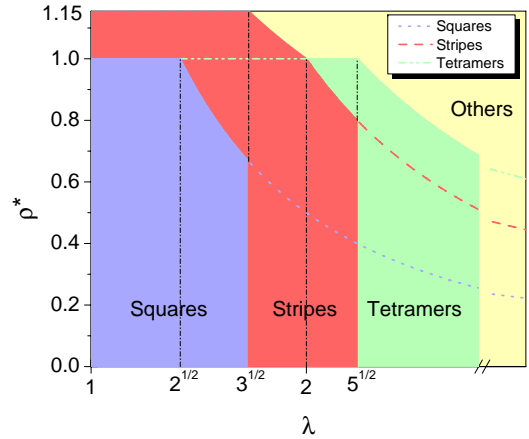}
\caption{Zero-temperature phase diagram. This figure shows the different
regions of the parameter space where different patterns appear. The blue zone
represents the region where squares are present; the red zone corresponds to
the appearance of stripes; finally, tetrameres appear in the green zone. The
yellow zone at the upper-right corner of the figure corresponds to other
structures not studied in this work. The blue (dotted), red (dashed) and green
(dash-dot-dot) define the maximum density at which squares, stripes, and
tetrameres could also appear, respectively, at finite temperatures.}%
\label{fig.3}%
\end{figure}

\begin{figure}[ptb]
\includegraphics[width=8.5cm]{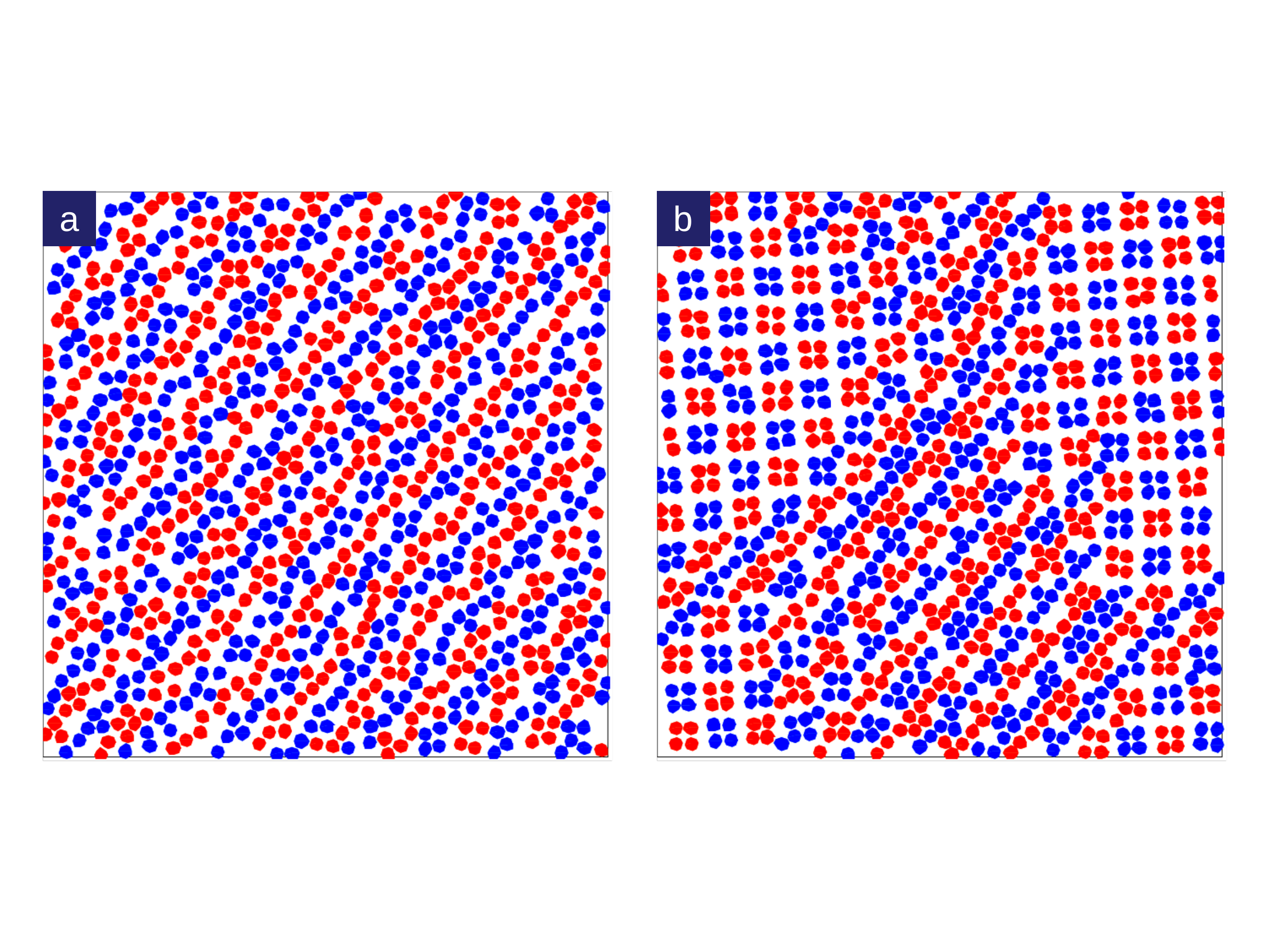}
\caption{Spatial configurations at temperatures $T^{\ast}=0.5$ and $T^{\ast}=0.1$. Snapshots taken at two temperatures along the annealing isochore. At high temperatures the system self-assemblies forming fluid-like disordered
stripes (a), while at low temperatures the system crystallizes in an ordered
square array of tetramers (b).}%
\label{fig.4}%
\end{figure}

\begin{figure}[ptb]
\includegraphics[width=8.5cm]{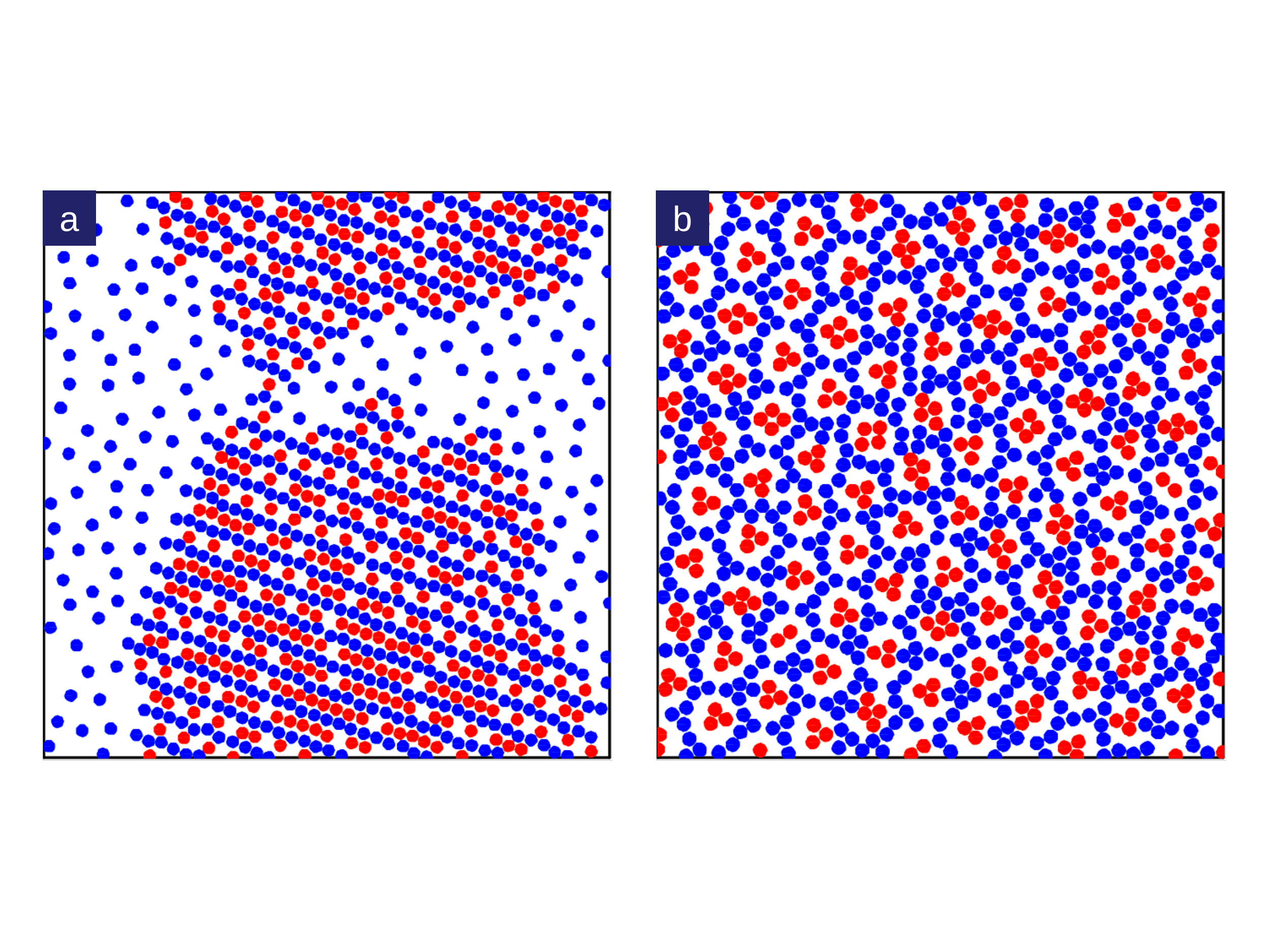}
\caption{Two representative configurations for asymmetric mixtures ($x=2$ and
$\rho^{\ast}=0.5$). In (a) we observe a phase separation with a region
consisting of a high-density blend of the two species and a region consisting
of a low-density pure phase of only the majority component, when $\lambda
=1.5$. In (b) we observe the formation of colloidal corrals, where the
majority component frames clusters of the minority component when $\lambda=2$.}%
\label{fig.5}%
\end{figure}

\end{document}